\documentclass[pra,twocolumn,showpacs,amsmath,amssymb,eqsecnum,superscriptaddress]{revtex4-1}
\usepackage{graphicx,bm,color,mathptmx,hyperref}

\newcommand{\tr}{\mathop{\mathrm{tr}}\nolimits}
\newcommand{\Tr}{\mathop{\mathrm{Tr}}\nolimits}
\newcommand{\Gal}[1]{\mathbb{F}_{#1}}

\begin{document}

\title{Optimal quantum tomography of  permutationally invariant qubits}
\author{A.~B.~Klimov}
\affiliation{Departamento de F\'{\i}sica, Universidad de Guadalajara,
 44420~Guadalajara, Jalisco, Mexico}

\author{G.~Bj\"{o}rk}
\affiliation{Department of Applied Physics, Royal Institute of Technology (KTH),
AlbaNova University Center, SE-106 91 Stockholm, Sweden}

\author{L. L. S\'anchez-Soto}
\affiliation{Departamento de \'{O}ptica, Facultad de F\'{\i}sica,
Universidad Complutense, 28040~Madrid, Spain}


\begin{abstract}
  Mutually unbiased bases determine an optimal set of measurements to
  extract complete information about the quantum state of a
  system. However, quite often \textit{a priori} information about the
  state exist, making some of the measurement bases superfluous. This is,
  for example, the case when multiqubit states belong to the permutationally
  invariant subspace.  In this paper we derive the minimal sets of
  mutually unbiased bases needed to tomographically reconstruct such
  states.
\end{abstract}

\maketitle

\section{Introduction}

The quantum state is a mathematical entity that encodes complete
information about a system: once it is known, the probabilities of the
outcomes of any measurement can be predicted. It seems thus
indisputable that ascertaining an unknown state accurately turns out
to be of uttermost importance for modern quantum technologies.
Broadly speaking, this is the scope of quantum
tomography~\cite{lnp:2004uq} which, over the past years, has evolved
from the initial theoretical~\cite{Vogel:1989zr} and experimental
concepts~\cite{Smithey:1993ly} to a widely acknowledged and fairly
standard method used  extensively for both
discrete~\cite{James:2001fk,*Thew:2002pd} and
continuous~\cite{Lvovsky:2009ys} variables.

However, the tomographic task becomes harder as we explore more
intricate systems. For example, for the simple case of $n$ qubits,
$2^{2n} - 1$ real numbers are required for its complete
characterization, while any von Neumann measurement gives only $2^n-1$
independent data.  Consequently, one will have to make at least
$2^n+1$ different such measurements before one can claim to know
everything about an \textit{a priori} unknown system.  With such a
scaling, it is clear that the methods rapidly become intractable for
present state-of-the-art experiments.

As a result, more sophisticated tomographical techniques are
called for. New protocols try to exploit the idea that the scheme is
explicitly optimized only for a particular kind of
states~\cite{Gross:2010dq,*Cramer:2010oq,*Shabani:2011hc,*Liu:2012kl}.
In that perspective, we look here at the specific but not unimportant example
of $n$ qubits prepared in an arbitrary state that is, however, known
to be invariant with  respect to any qubit permutation. This may be
due to, e.g., a permutationally invariant preparation Hamiltonian. In
this instance, the associated Hilbert space has dimension $n+1$, and
therefore it should be possible to reconstruct such a state with only
$n+2$ von Neumann measurements.

Permutationally invariant qubit states are employed in diverse quantum
information strategies~\cite{Stockton:2003fv,*Bartlett:2003dz,
  *Cabello:2007ij,*Fiurasek:2009bs,*Hentschel:2011fu}.  They are also
optimal for quantum metrology~\cite{Berry:2000qa,*Demkowiczi:2009kl}
and play an important role in the characterization of locally
non-interconvertible entanglement classes~\cite{Rajagopal:2002mi,
*Devi:2007pi,*Usha:2007lh,*Toth:2009uq,*Markham:2011fk,*Ribeiro:2011kx,
*Lyons:2011bh,*Wang:2012uq,*Devi:2012fu}. Through all this paper we
take permutational invariance for granted; theoretical tests of this
property (others than full tomography) have been put 
forward in~\cite{Majewski:2009kx,*Hillery:2011vn}.

Recently, a number of suggestions have appeared for an
efficient generation of different entangled permutationally symmetric
qubit states~\cite{Yamamoto:2002vn,*Kiesel:2007ys,
  *Wieczorek:2009zr,*Bastin:2009fk,*Prevedel:2009ly}. The tomography
of such states has already been discussed in
Ref.~\cite{Ariano:2003qf}, and a four-qubit experiment has 
been performed~\cite{Toth:2010dq}. However, in these proposals the
measurements have been chosen as a set of informationally complete
projectors. This may provide a certain experimental simplicity (e.g.,
for spin states it may be possible to simply use the orientation of a
Stern-Gerlach apparatus to chose the projector), but is by no means an
optimal strategy.

The number of separate von Neumann measurements needed for a complete
state determination is optimal when the bases in which those
measurements are performed are mutually unbiased~\cite{Durt:2010cr}
(in the standard $n$ qubit Pauli tomography, $3^{n}$ different
settings are needed, while in this optimal approach, $2^{n}+1$ are
enough). In fact, the notion of mutually unbiased bases (MUBs) emerged
in the seminal work of Schwinger~\cite{Schwinger:1960dq,
  *Schwinger:1960cr, *Schwinger:1960nx} and has turned into a
cornerstone of quantum information, mainly due to the spotlight placed
on them by the elegant work of Wootters and
coworkers~\cite{Wootters:1987ve,*Wootters:1989qf,*Wootters:2004qf,
  *Gibbons:2004cr,*Wootters:2006nx}.  MUBs are endowed with the
property of being maximally incompatible, in the sense that a state
giving precise results in one set (i.e., one of the basis states)
produces maximally random results when measured in another basis in
the MUB set.

Another remarkable advantage of the MUB-based tomography is that each
measured probability determines a single element of the density matrix
so, in principle, there should be no need for a numerical data
inversion to reconstruct the corresponding state.  In practice,
however, experimental noise and measurement imperfections may yield an
unphysical density matrix, so a fitting procedure might still be
needed.
 
For all these compelling reasons, we think it is worthwhile to first prove that
minimal complete sets of MUBs exist for the tomography of a permutationally
invariant $n$-qubit state, and subsequently  show how to construct
them.  This is precisely the goal of this paper. Of course, these
advantages come with a price: those minimal  MUB sets are, in general,
comprised of entangled projectors, which  renders their experimental
implementation more challenging than the product Pauli projectors.

\section{Mutually unbiased measurements for qubits}

A compact way of labeling $n$-qubit states consists in using the
finite field $\mathbb{F}_{2^{n}}$ (the reader interested in
mathematical details is referred, e.g., to the excellent monograph by
Lidl and Niederreiter~\cite{Lidl:1986uq}). For our purposes, this can
be considered as a linear space spanned by an abstract basis $\{
\theta_{1}, \ldots ,\theta_{n} \} $, so that given a field element
$\nu$ (henceforth, they will be denoted by Greek letters) the
expansion $\nu =\sum_{i=1}^{n} n_{i} \, \theta_{i}$ (with $n_{i} \in
\mathbb{Z}_{2}$) allows us the identification $ \nu \Leftrightarrow (
n_{1}, n_{2}, \dots , n_{n} )$.

Moreover, the basis can be chosen to be orthonormal with respect to
the trace operation (the self-dual basis); that is, $\tr ( \theta_{i}
\theta_{j}) =\delta_{ij}$, where $\tr ( \nu ) = \nu + \nu^{2} +
\ldots +\nu^{2^{n-1}}$ and maps $\mathbb{F}_{2^{n}}$ onto the base
field $\mathbb{Z}_{2}$. In this way, to each qubit we associate a
particular element of the self-dual basis: $i$th qubit $\Leftrightarrow$
$\theta_{i}$.

Let $\{ | \nu \rangle \} $ be an orthonormal basis in the Hilbert
space of the system, which is isomorphic to
$\mathbb{C}^{2^{n}}$. Operationally, the elements of the basis are
labeled by powers of a primitive element. These vectors are
eigenvectors of the operators $Z_{\alpha}$ belonging to the
generalized Pauli group~\cite{Chuang:2000fk}, whose basic generators
are
\begin{equation}
  Z_{\alpha} =\sum_{\nu} (-1)^{\tr (\nu \alpha)} \,
  |\nu \rangle \langle \nu | \, ,
  \qquad
  X_{\beta} = \sum_{\nu} |\nu +\beta \rangle \langle \nu | \,,
  \label{zx}
\end{equation}
with $\alpha, \beta \in \mathbb{F}_{2^{n}}$. Notice that
in the self-dual basis these operators factorize as
\begin{equation}
  Z_{\alpha} = \sigma_{z}^{a_{1}} \otimes \cdots
  \otimes  \sigma_{z}^{a_{n}} \, ,
  \qquad
  X_{\beta}=\sigma_{x}^{b_{1}}\otimes \cdots
  \otimes \sigma_{x}^{b_{n}} \, ,
  \label{zxfac}
\end{equation}
where $a_{i} = \tr ( \alpha \theta_{i})$ and $b_{i}= \tr (\beta
\theta_{i})$ are the corresponding expansion coefficients for $\alpha
$ and $\beta $ in that basis. The single-qubit Pauli operators
$\sigma_{z}$ and $\sigma_{x}$ can be expressed in  the standard basis
of the two-dimensional Hilbert space $\mathbb{C}^{2}$ as
\begin{equation}
  \sigma_{z} = | 1 \rangle \langle 1 | - |0 \rangle \langle 0 |,
  \qquad
  \sigma_{x} = | 0 \rangle \langle 1 | + | 1 \rangle \langle 0 | \, .
  \label{sigmas}
\end{equation}

In addition, we have the commutation relation
\begin{equation}
  Z_{\alpha}X_{\beta} = (-1)^{\tr (\alpha \beta)} \,
  X_{\beta} Z_{\alpha} \, .
\end{equation}
This is the discrete counterpart of the Heisenberg-Weyl group for
continuous variables and  the hallmark of noncommutativity.  Moreover, 
$X_{\alpha}$ and $Z_{\alpha}$ are related through the finite Fourier
transform~\cite{Vourdas:2003sa,*Vourdas:2004,*Vourdas:2005,*Vourdas:2007}
\begin{equation}
  \mathcal{F}=\frac{1}{\sqrt{2^{n}}} \sum_{\nu  ,\nu^{\prime}}
  (-1)^{\tr (\nu \nu^{\prime})} \,
  |\nu  \rangle \langle \nu^{\prime}| \, ,
  \label{FTcomp}
\end{equation}
so that $X_{\alpha}=\mathcal{F} \, Z_{\alpha}
\,\mathcal{F}$~\cite{Klimov:2005oq}.

We next recall~\cite{Gibbons:2004cr} that the grid specifying the
phase space for $n$ qubits can be appropriately labeled by the
discrete points $(\alpha, \beta)$, which are precisely the indices of
the operators $Z_{\alpha}$ and $X_{\beta}$: $\alpha$ is the
``horizontal'' axis and $\beta$ the ``vertical'' one.  On this grid
one can introduce a variety of geometrical
structures with much the same properties as in the
continuous case~\cite{Klimov:2007bh,*Klimov:2009bk,*Klimov:2012kl};
the  simplest are the straight lines passing through the origin
(also called rays). These rays have a quite remarkable property: the
monomials $\{ Z_{\alpha} X_{\mu \alpha} \}$ labeled by points of the
same ray commute with each other, and thus have a common system of
eigenvectors, which we shall label as $ | \nu , \mu \rangle$. Without
going into details, they can be constructed as
\begin{equation}
  |\nu ,\mu \rangle  =  X_{\nu} V_{\mu}
  |0\rangle \otimes \cdots \otimes |0\rangle \,  ,
\label{eq:slope}
\end{equation}
where $ V_{\mu}$ is a finite rotation that changes the slope $\mu$ of
the rays. Therefore, states with the same slope index $\mu$ span an
orthogonal basis.  The explicit construction of $ V_{\mu}$ can be found in
Ref.~\cite{Klimov:2007bh,*Klimov:2009bk,*Klimov:2012kl}. In
this way, both the state index $\nu $ and the slope (i.e., basis) $\mu$ run
over the $2^{n}$ elements of $\Gal{2^{n}}$.  Of course, there is an
extra basis of ``infinite'' slope (corresponding to the ``vertical''
axis) that cannot be obtained through a rotation:
\begin{equation}
  |\tilde{\nu}\rangle = X_{\nu}  \mathcal{F}  |0\rangle \otimes \cdots
    \otimes |0\rangle  \, .
\label{eq:vert}
  \end{equation}
  
  Let us enumerate for the time being these vectors by $|\nu , k \rangle$,
  where $k$ runs the $2^{n}$ values of $\mu$ in (\ref{eq:slope}) plus
  the extra basis in (\ref{eq:vert}). One can easily check that
  \begin{equation}
    |\langle \nu ,k|\nu ^{\prime}, k^{\prime} \rangle |^{2}=
    \delta_{kk^{\prime}} \delta_{\nu \nu^{\prime}} +
    \frac{1}{2^{n}} (1-\delta_{kk^{\prime}}) \, ,
  \end{equation}
  so they constitute a set of MUBs.  In other words, the complete set
  of ($2^{n}+1$) mutually unbiased projectors
  \begin{equation}
    \label{eq:MUBproj}
    \mathcal{P}_{\nu ,k}=|\nu ,k\rangle \langle \nu ,k| \, ,
  \end{equation}
  defines a complete scheme, in the sense that the measured
  probabilities
  \begin{equation}
\label{eq:normnk}
    p_{\nu, k} = \Tr ( \varrho \mathcal{P}_{\nu,k} ) \, ,
    \qquad
    \sum_{\nu}  p_{\nu, k} = 1 \, ,
  \end{equation}
  determine completely the density matrix through
  \begin{equation}
    \varrho + \openone=\sum_{k=1}^{2^{n}+1} \sum_{\nu} p_{\nu,k}
    \mathcal{P}_{\nu ,k} \,  .
    \label{Eq:rho}
  \end{equation}
  Here, we have used $\Tr$ (with capital T) for the standard trace in
  Hilbert space. Note that the structure of this MUB set is preserved
  under any local unitary transformation, so any factorizable,
  complete basis can be chosen as a computational basis.

\section{Optimal measurement scheme under permutational symmetry}

As heralded in the Introduction, we wish to design a minimal
MUB tomographical scheme for systems that remain invariant under
all possible interchanges of its different particles. This invariance
could be simply stated as
\begin{equation}
  \Pi_{pq} \, \varrho \, \Pi_{pq} = \varrho \, ,
  \label{Eq:permutation}
\end{equation}
where $p \neq q, \;p, q\in \{1, \ldots ,n\}$. The elements $\Pi_{pq}$
of the permutation group are known as swap operators, as they exchange
the states of the $p$-th and the $q$-th qubits; i. e.,
\begin{equation}
  \Pi_{pq} |\ldots ,a_{p},\ldots ,a_{q},\ldots \rangle
  =|\ldots ,a_{q},\ldots ,a_{p},\ldots \rangle \,.
\end{equation} 

It has recently been shown~\cite{Moroder:2012uq} that any permutationally
invariant  $n$-qubit state defined via Eq.~(\ref{Eq:permutation}) can be written as
\begin{equation}
  \varrho_{\mathrm{PI}} = \bigoplus_{j=j_{\min}}^{n/2} p_{j} \;
  \varrho_{j} \otimes R_{j} \, .
  \label{rho decomp}
\end{equation}
The summation runs over different total spin numbers starting from
$j_{\min} \in \{0, 1/2 \}$, depending on whether $n$ is even or odd. $\varrho_{j}$
is the density matrix of a $j$-spin state and $p_{j}$ are the
associated probabilities. In addition, $R_{j} = \openone/ \dim
\mathcal{K}_{j} $ and the factor $\dim \mathcal{K}_{j} $ comes
from the degeneracy of the subspaces appearing in the decomposition
of the total Hilbert space $\mathcal{H} = \mathbb{C}^{{2}^{n}}$ in the
form
\begin{equation}
\mathcal{H} = \bigoplus_{j=j_{\min}}^{n/2} \mathcal{H}_{j}
\otimes \mathcal{K}_{j} \, .
\label{decompos}
\end{equation}
Here, $\mathcal{H}_{j}$ is a spin Hilbert space of dimension $\dim
\mathcal{H}_{j}=2j+1$, while $\mathcal{K}_{j}$ are referred as
multiplicative spaces that account for the different possibilities to
obtain a spin $j$. One can show that~\cite{Varshalovich:1988ct}
\begin{equation}
  \label{eq:mlt}
  \dim \mathcal{K}_{j} =
\left (
\begin{array}{c}
n \\
n/2 - j
\end{array}
\right )
-
\left (
\begin{array}{c}
n \\
n/2 - j -1
\end{array}
\right ) \, .
\end{equation}
This means that a permutationally invariant density operator only
contains nontrivial parts of the spin Hilbert spaces, and  there
are no coherences between different spin states. In other words, any
of these states can be parsed into a block-diagonal form that   has been
exploited in several contexts~\cite{Adamson:2007uq,*Shalm:2009vn,
*Karassiov:2005ss,*Marquardt:2007bh,*Muller:2012ys}.

The crucial observation for what follows is that, at the level
of field elements, the action of the permutation operator $\Pi_{pq}$
on a state is tantamount of
\begin{equation}
  \mu \mapsto  \mu +\varepsilon \tr (\mu \varepsilon ) \, ,
\end{equation}
where $\varepsilon = \theta_{p}+\theta_{q}$, with $\theta_{p}$
($\theta_{q}$) being the $p$-th ($q$-th) element of the self-dual
basis. Since the field element addition is commutative, the operator is
symmetric in $p$ and $q$, as it should. In algebraic terms, we have
then
\begin{equation}
  \Pi_{pq} = \sum_{\kappa}|\kappa +\varepsilon
  \tr ( \varepsilon \kappa ) \rangle \langle \kappa | \, .
  \label{P}
\end{equation}
Using this field representation, we can check that the mutually
unbiased projectors (\ref{eq:MUBproj}) transform as
\begin{equation}
  \Pi_{pq} \, \mathcal{P}_{\nu, \mu} \, \Pi_{pq}=
  \mathcal{P}_{\nu +\varepsilon \tr (\nu\varepsilon),
    \mu +\varepsilon  \tr(\nu \varepsilon )} \, ,
\end{equation}
whence the density matrix in the tomographic
representation~(\ref{Eq:rho}) is transformed as
\begin{eqnarray}
  \Pi_{pq} \, (\varrho + \openone ) \, \Pi_{pq} & = & 
 \sum_{\mu ,\nu} p_{\nu ,\mu}  
 \mathcal{P}_{\nu +\varepsilon \tr (\nu\varepsilon),   
  \mu +\varepsilon  \tr(\nu \varepsilon )}  \nonumber \\
  & + & \sum_{\nu}\tilde{p}_{\nu}
  \tilde{\mathcal{P}}_{\nu +\varepsilon \tr(\nu \varepsilon )} \, .
\label{eq:coninv}
\end{eqnarray}
The last term is just the contribution from the basis
(\ref{eq:vert}), which we split for notational simplicity.

If we perform the change of variables
\begin{eqnarray}
  \nu ^{\prime} =\nu +\varepsilon \tr (\nu \varepsilon ),
  \qquad
  \mu ^{\prime} =\mu +\varepsilon \tr ( \mu \varepsilon ),
  \label{numu1}
\end{eqnarray}
we can recast (\ref{eq:coninv}) in the form
\begin{eqnarray}
  \Pi_{pq} \, (\varrho +  \openone ) \Pi_{pq} & = &
  \sum_{\mu ^{\prime},\nu ^{\prime}}
  p_{\nu^{\prime}+\varepsilon \tr(\nu^{\prime} \varepsilon ),
    \mu^{\prime}+\varepsilon \tr ( \mu^{\prime} \varepsilon)}
  \mathcal{P}_{\nu^{\prime}, \mu^{\prime}} \nonumber \\
  & + & \sum_{\nu^{\prime}}\tilde{p}_{\nu ^{\prime}+
    \varepsilon \tr (\nu^{\prime} \varepsilon)}
  \tilde{\mathcal{P}}_{\nu^{\prime}}.
\end{eqnarray}
Consequently, the invariance condition (\ref{Eq:permutation}) leads to the
following restrictions on the measured probabilities:
\begin{equation}
  p_{\nu +\varepsilon \tr (\nu \varepsilon),
    \mu +\varepsilon \tr ( \mu \varepsilon )} =
  p_{\nu , \mu} \, , \qquad \forall \varepsilon \, .
  \label{Eq:condition}
\end{equation}
Obviously, the probabilities $p_{\nu ,\mu}$ should be also invariant
under all consecutive index permutations.

\section{Physical discussion}

The above basic expression can be given a transparent physical
meaning.  Indeed, let us expand $\nu$ and $\mu $ in the self-dual
basis
\begin{equation}
  \mu =\sum_{i=0}^{n}m_{i}  \theta_{i}  \, ,
  \qquad
  \nu =\sum_{i=0}^{n} n_{i}  \theta_{i}  \, ,
\end{equation}
with $m_{i}, n_{i} \in \mathbb{Z}_{2}$ and analogous expansions for
the transformed indices in (\ref{numu1}).  For a given $\varepsilon =
\theta_{p} +\theta_{q}$, one can check that $m_{i}^{\prime}=m_{i}$,
except for $m_{p}^{\prime}=m_{q}$ and $m_{q}^{\prime}=m_{p}$, and
similarly for $n_{i}^{\prime}$. That is, a change of the index $\nu $
of the states in a basis simply results in a reshuffling of its
states. Therefore, such transformations do not give any new
tomographic projectors for a permutationally invariant state.

The transformation of $\mu $ implies that measurements by MUBs
corresponding to $\mu $ indices with the same number of non-zero
components in the self-dual basis [the length of the word $|\mu |$
corresponding to the binary string $(a_{0},a_{1,}...,a_{n})$] give the
same information. In short, the projectors labeled by, e.g.,
$\mu=(1,1,0,\ldots,0)$ and $\mu^{\prime} =(0,\ldots,0,1,1)$ are
equivalent. The computational basis, associated with $\mu =0$,
automatically satisfies (\ref{Eq:condition}) for all $\nu$. Similarly,
the $X$ basis also satisfies (\ref{Eq:condition}) because it has no
$\mu $ dependence ($[\Pi _{pq}, \mathcal{F} ]=0$). Therefore, these
two bases remain invariant under any qubit permutation. This allows us
to count the total number of measurements needed for a complete
reconstruction of the density matrix, which is just $ n+2$. This
result could be expected, for the Hilbert space dimension of the
permutation invariant system is $n+1$.

Since the permutation group acts simultaneously on both indices $\mu $
and $\nu$, there are different orbits of equivalent probabilities that
are defined not only by the length $|\mu |$ but also by the mutual
symmetry properties of the indices representing the number of the
basis and the element in each basis. In particular, for $\mu \neq 0$
each orbit representative is labeled by three lengths $m=|\mu |,l=|\nu
|,s=|\mu +\nu |$, i.e. $p_{\nu ,\mu} = p (m, l, s)$.  For the
computational and the Fourier bases the orbits are characterized only
by $|\nu |$; for instance, $p_{\nu, 0}=p_{0} ( l )$.  Accordingly, in
each basis not all the probabilities should be measured, but only
those that belong to different orbits, which leads to a reduction of
the experimental errors.  Since for given $m$ and $l$, $s$ runs from
$|m - l|$ to $\min(m+l, 2n - m - l, n)$ in steps of two, the number of
orbits turns out to be $1+n (n^{2} +6 n+17 )/6$. Bearing in mind the
normalization condition (\ref{eq:normnk}), we find that there are $n (
n^{2} +6 n+17)/6$ independent probabilities $p(m, l, s)$, which
completely define the density matrices appearing in the
decomposition~(\ref{rho decomp}). Projectors corresponding to the same
probabilities are given by the condition~(\ref{Eq:condition}).

The final reconstruction takes the form
\begin{eqnarray}
\varrho + \openone  &=& \sum_{m=0}^{n} \sum_{l=1}^{n}
\sum_{s= | m-l }^{\min ( m+l,n )}  p ( m, l, s)
\sum_{\mu} \sum_{\nu}  \mathcal{P}_{\nu (l,s) , \mu (m,s)}  \nonumber \\
& + &\sum_{l=1}^{n} p_{0} ( l ) \sum_{\nu} \mathcal{P}_{\nu (l), 0}  +
\sum_{l=0}^{n} \tilde{p}(l) \sum_{\nu} \tilde{\mathcal{P}}_{\nu (l)}
+ \sum_{\mu \nu } p_{0, \mu } \mathcal{P}_{0 , \mu }  \nonumber \\
\label{TR}
\end{eqnarray}
where the sum on $\mu $ and $\nu $ run over all the field elements
such that $|\mu (m,s)|=m, l=|\nu (l,s)|$, $s=|\mu +\nu |$, and
$p_{0, \mu}= 1 - \sum_{\nu} p_{\nu ,\mu}$, and
$\tilde{p}_{0}=1-\sum_{\nu}\tilde{p}_{\nu}$.

For instance, for two qubits, the field $\Gal{2^{2}}$ has the
primitive element defined by the irreducible polynomial $\theta
^{2}+\theta +1 = 0$. Therefore, $\theta_{1}=\theta $ and
$\theta_{2}=\theta ^{2}$, so that $\theta^{3}=\theta \theta^{2} =
\theta (1+\theta )= \theta_{1}+\theta_{2}$. In this case, only
measurements in the bases with $\mu =\theta_{1}$ (or $\mu
=\theta_{2}$) and $\mu =\theta_{1}+\theta_{2}$ (apart from measurements
in the computational and $X$ bases) are required. The 9 independent
measured probabilities [$p_{\theta_{1},0}$ and
$p_{\theta_{1}+\theta_{2},0}$, from the $Z$ basis, $p_{\theta_{1},
  \theta_{1}}$, $p_{\theta_{2}, \theta_{1}}$, and
$p_{\theta_{1}+\theta_{2}, \theta_{1}}$ from basis 1,
$p_{\theta_{1},\theta_{1}+\theta_{2}}$ and $p_{\theta_{1}+\theta_{2},
  \theta_{1}+\theta_{2}}$ from basis 3, and $\tilde{p}_{\theta _{1}}$
and $\tilde{p}_{\theta_{1}+\theta_{2}}$ from the $X$ basis] are
representatives of the equivalent probabilities orbits. This selection
gives an explicit reconstruction form that reads as 
\begin{widetext}
  \begin{eqnarray}
    \varrho + \openone &=& p_{\theta_{1},0} \left (
      \mathcal{P}_{\theta_{1}, 0} +
      \mathcal{P}_{\theta_{2},  0} \right ) +
    p_{\theta_{1}+ \theta_{2},0}
    \mathcal{P}_{\theta_{1}+ \theta_{2},0}  \nonumber  \\ 
    & + & p_{\theta_{1}, \theta_{1}} \left (
      \mathcal{P}_{\theta_{1}, \theta_{1}} +
      \mathcal{P}_{\theta_{2},\theta_{2}} \right)
    +  p_{\theta_{2},\theta_{1}} \left (
      \mathcal{P}_{\theta _{2}, \theta_{1}}+
      \mathcal{P}_{\theta_{1}, \theta_{2}} \right ) +  p_{\theta_{1}+\theta_{2},\theta_{1}}
    \left (\mathcal{P}_{\theta_{1}+\theta _{2}, \theta_{1}} + 
      \mathcal{P}_{\theta_{1}+\theta_{2}, \theta_{2}} \right ) \nonumber \\
    & + & p_{\theta_{1}, \theta_{1}+\theta_{2}} 
    \left ( \mathcal{P}_{\theta_{1}, \theta_{1}+\theta_{2}} +
      \mathcal{P}_{\theta_{2}, \theta_{1}+\theta_{2}} \right)  + 
    p_{\theta_{1}+\theta_{2}, \theta_{1}+\theta_{2}}
    \mathcal{P}_{\theta_{1}+\theta_{2}, \theta_{1}+\theta_{2}}  \nonumber \\
    & + &\tilde{p}_{\theta_{1}} \left (
      \tilde{\mathcal{P}}_{\theta_{1}}
      + \tilde{\mathcal{P}}_{\theta_{2}}
    \right)  +
    \tilde{p}_{\theta_{1}+\theta_{2}}
    \tilde{\mathcal{P}}_{\theta_{1}+\theta_{2}} \nonumber \\
    & + &  p_{0,0} \mathcal{P}_{0, 0} + p_{0, \theta_{1}} \mathcal{P}_{0, \theta_{1}} +
    p_{0, \theta_{2}}  \mathcal{P}_{0, \theta_{2}} +
    p_{0, \theta_{1}+\theta_{2}} \mathcal{P}_{0, \theta_{1}+\theta _{2}}
    + \tilde{p}_{0} \tilde{\mathcal{P}}_{0} \, ,
  \end{eqnarray}
\end{widetext}
where $p_{0, \mu}= 1- \sum_{\nu} p_{\nu ,\mu}$ and thus can be derived
from the nine independent, measured probabilities. Similarly,
$\tilde{p}_{0}=1-2 \tilde{p}_{\theta_{1}}-\tilde{p}_{\theta_{1}+\theta_{2}}$.

For this problem the computational basis is
\begin{equation}
  \{X_{\nu} |0\rangle \}= \{|\nu \rangle \} =
  \left\{
    \left(
      \begin{array}{c}
        1 \\
        0 \\
        0 \\
        0
      \end{array}
    \right )  ,
    \left (
      \begin{array}{c}
        0 \\
        1 \\
        0 \\
        0
      \end{array}
    \right )  ,
    \left (
      \begin{array}{c}
        0 \\
        0 \\
        1 \\
        0
      \end{array}
    \right )  ,
    \left (
      \begin{array}{c}
        0 \\
        0 \\
        0 \\
        1
      \end{array}
    \right )
  \right \} \, .
\end{equation}
The three remaining bases (apart from a normalization factor) are
\begin{eqnarray}
  & \{ X_{\nu}|\theta_{1}\rangle \}  = \left\{
    \left (
      \begin{array}{c}
        1 \\
        i \\
        1 \\
        -i
      \end{array}
    \right ) , \left (
      \begin{array}{c}
        i \\
        1 \\
        -i \\
        1
      \end{array}
    \right ) ,\left (
      \begin{array}{c}
        1 \\
        -i \\
        1 \\
        i
      \end{array}
    \right ) ,\left (
      \begin{array}{c}
        -i \\
        1 \\
        i \\
        1
      \end{array}
    \right ) \right \} , &  \nonumber \\
  & \{X_{\nu}|\theta_{2}\rangle \} =\left \{
    \left (
      \begin{array}{c}
        1 \\
        1 \\
        i \\
        -i
      \end{array}
    \right ) ,
    \left (
      \begin{array}{c}
        1 \\
        1 \\
        -i \\
        i
      \end{array}
    \right ) ,
    \left (
      \begin{array}{c}
        i \\
        -i \\
        1 \\
        1
      \end{array}
    \right ) ,
    \left (
      \begin{array}{c}
        -i \\
        i \\
        1 \\
        1
      \end{array}
    \right ) \right\} , & \\
  & \{X_{\nu}|\theta_{1}+\theta_{2}\rangle
  \} =  \left\{ \left (
      \begin{array}{c}
        i \\
        1 \\
        1 \\
        -i
      \end{array}
    \right)  , \left (
      \begin{array}{c}
        1 \\
        i \\
        -i \\
        1
      \end{array}
    \right ) ,
    \left (
      \begin{array}{c}
        1 \\
        -i \\
        i \\
        1
      \end{array}
    \right ) ,
    \left (
      \begin{array}{c}
        -i \\
        1 \\
        1 \\
        i
      \end{array}
    \right ) \right\}  \, , & \nonumber
\end{eqnarray}
while the one corresponding to (\ref{eq:vert}) turns out to be
\begin{equation}
      \{|\tilde{\nu}\rangle \}=\left\{ \left (
          \begin{array}{c}
            1 \\
            1 \\
            1 \\
            1
          \end{array}
        \right ) ,\left (
          \begin{array}{c}
            1 \\
            -1 \\
            1 \\
            -1
          \end{array}
        \right ) \left (
          \begin{array}{c}
            1 \\
            1 \\
            -1 \\
            -1
          \end{array}
        \right ) , \left (
          \begin{array}{c}
            1 \\
            -1 \\
            -1 \\
            1
          \end{array}
        \right ) \right\} .
    \end{equation}

%%%%%%%%%%%%%%%%%%%%%%%%%%%%%%%%%%%%%%%%%%%%%
    \begin{table}[b]
\squeezetable
      \caption{\label{tab:table1}
        Allowed values of $m$, $l$ and $s$ for the 24 independent
        orbits in the three-qubit case. The tilde indicates that the
        corresponding probabilities are measured in the $X$ basis. The last row (denoted
        $\#$) gives  the number of (equivalent) probabilities in each orbit. }
      \begin{ruledtabular}
        \begin{tabular}{lcccccccccccccccccccccccc}
          $m$ & 0 & 0 & 0 & 0 & $\tilde{0}$ &  $\tilde{0}$ &  $\tilde{0}$ &
          $\tilde{0}$ & 1 &  1 & 1 &  1 & 1 &  1 & 2 & 2 &  2 & 2 & 2 & 2 & 3 &
          3  & 3 & 3 \\
          $l$ & 0 & 1 & 2 & 3 & $\tilde{0}$ &  $\tilde{1}$ &  $\tilde{2}$ &
          $\tilde{3}$ & 0 &  1 & 1 &  2 & 2 &  3 & 0 & 1 &  1 & 2 & 2 & 3 & 0 &
          1  & 2 & 3 \\
          $s$ & 0 & 1 & 2 & 3 & $\tilde{0}$ &  $\tilde{1}$ &  $\tilde{2}$ &
          $\tilde{3}$ & 1 &  0 & 2 &  1 & 3 &  2 & 2 & 1 &  3 & 0 & 3 & 1 & 3 &
          2  & 1 & 0 \\
          $\#$ & 1 & 3 & 3 & 1 & 1 &  3 &  3 & 1 & 3 &  3 & 6 &  6 & 3
          &  3 & 3 & 6 &  3 & 3 & 6 & 3 & 1 & 3  & 3 & 1 \\
        \end{tabular}
      \end{ruledtabular}
    \end{table}
%%%%%%%%%%%%%%%%%%%%%%%%%%%%%%%%%%%%%%%%%%%%%%%%%%%%%%%%%%

    Of these five MUBs, only, e.g., $\{|\nu \rangle \}$, $
    \{X_{\nu}|\theta_{1}\rangle \}$, $\{X_{\nu}|\theta_{1}+\theta
    _{2}\rangle \}$, and the $\{|\tilde{\nu}\rangle \}$ are needed to
    tomographically reconstruct a permutationally invariant two-qubit
    state. If we permute the second and the third qubit (and the state
    to be tomographed would not change due to such permutation) it is
    readily seen that the permuted basis $\{X_{\nu}|\theta_{2}\rangle
    \}$ becomes the non-permuted basis $\{X_{\nu}|\theta_{1}\rangle
    \}$ (but with the middle two vectors interchanged). Hence, the two
    bases extract identical information from the state, and hence one
    of them can be disregarded. They both have one nonzero component
    in the self dual basis and are therefore directly related by a
    permutation as shown by (\ref{Eq:condition}).

Before we conclude, let us briefly address the case of three qubits.  In
Table~\ref{tab:table1} we give the values of $l, m, s$ for the 24
independent orbits (all in all, we get 72 probabilities).
 Taking into account that 5 probabilities (each one defining an orbit)
 can be determined from the normalization condition (\ref{eq:normnk})
 [for example, we can fix $p(0,0,0)$, $\tilde{p}( 0, 0, 0,)$,
 $p(1,0,1)$,  $p(2,0,2)$, and  $p(3,0,3)$], we	arrive at 19
orbits that  determine any symmetric density matrix.

\section{Conclusions}

We have developed a method to generate a  minimal set of MUBs 
needed to tomographically reconstruct a state
consisting of $n$ qubits, when the state is invariant under the
permutation of the qubits.  Such a state spans an $n+1$ dimensional
Hilbert space. Consequently the smallest set of bases one can hope to
use is $n+2$, and indeed our method provides a minimal set. 

MUBs are not strictly necessary to reconstruct such a state, but they
have the advantage of capturing maximally different
aspects of the state. Moreover, as the bases constitute complete sets
of states in the $2^n$-dimensional space of $n$ qubits, they can in
principle be implemented as von Neumann measurements and not as
individual projectors or positive operator valued measures. The price
is that the MUB projectors are for the most part highly entangled, so
their experimental implementation can be difficult.

\begin{acknowledgments}
  It is a pleasure acknowledging helpful discussions with Hubert de Guise.
  Financial support from the mexican CONACyT (Grant 106525), the
  Swedish Foundation for International Cooperation in Research and
  Higher Education (STINT), the Swedish Research Council (VR) through
  its Linn\ae us Center of Excellence ADOPT (contract 621-2011-4575),
  the Spanish DGI (Grant FIS2011-26786) and the UCM-BSCH program
  (Grant GR-920992). is acknowledged.
\end{acknowledgments}

%\bibliography{Permtomo}

%merlin.mbs 2010-03-15 4.21a (PWD, AO, DPC)

\end{document}